%% file: main.tex
\documentclass[manuscript,screen]{acmart}
   
\renewcommand\footnotetextcopyrightpermission[1]{}

\usepackage{tcolorbox}
\usepackage{xspace}
\usepackage{enumitem}
\usepackage{algorithm}
\usepackage{algorithmic}

\floatname{algorithm}{Function}
\usepackage{amsthm}
\theoremstyle{acmdefinition}

\usepackage{booktabs}   
\usepackage{multirow}   
\usepackage{tabularx}   
\usepackage{subcaption} 
\usepackage{diagbox}

\usepackage{graphicx}
\usepackage{booktabs}

\newcommand{\todo}[1]{\textcolor{black}{#1}}
\newcommand{\tool}{\textsc{Aladdin}\xspace}
\newcommand{\navbase}{\textsc{Prompt.based}\xspace}
\newcommand{\oraclebase}{\textsc{AugmenTest}\xspace}
\newcommand{\numfaulty}{64\xspace}
\newcommand{\numcorrect}{80\xspace}
\newcommand{\summary}[1]{
 \begin{center}
  \begin{tcolorbox}[colback=gray!10,colframe=black!25,width=1\columnwidth,arc=1mm, auto outer arc,boxrule=0.5pt,boxsep=3pt,left=1pt,right=1pt,top=0pt,bottom=0pt]
  \textbf{\textit{Summary}:} #1
  \end{tcolorbox}
 \end{center}
}

\begin{document}

\title{Automated Functional Testing for Malleable Mobile Application Driven from User Intent}

\author{Yuying Wang}
\email{2534044@tongji.edu.cn}
\author{Kaifeng Huang}
\authornote{Corresponding Authors}
\email{kaifengh@tongji.edu.cn}
\author{Hao Deng}
\authornotemark[1]
\email{denghao1984@tongji.edu.cn}
\author{Zhiyuan Sun}
\email{2433274@tongji.edu.cn}
\author{Jinxuan Zhou}
\email{2451751@tongji.edu.cn}
\author{Shengjie Zhao}
\email{shengjiezhao@tongji.edu.cn}
\affiliation{%
  \institution{School of Computer Science and Technology, Tongji University}
  \country{China}
}

\renewcommand{\shortauthors}{AAA et al.}

\input{src/ch00-abstract}

\maketitle

\input{src/ch01-introduction}
\input{src/ch02-motivation}

\input{src/ch03-approach}
\input{src/ch04-evaluation-setup}

\input{src/ch05-evaluation}

\input{src/ch06-related}

\input{src/ch07-conclusion}

\clearpage

\bibliographystyle{ACM-Reference-Format}
\bibliography{src/reference.bib}

\clearpage

\end{document}

%% file: src/ch00-abstract.tex
\begin{abstract}

Software malleability allows applications to be easily changed, configured, and adapted even after deployment. While prior work has explored configurable systems, adaptive recommender systems, and malleable GUIs, these approaches are often tailored to specific software and lack generalizability. In this work, we envision per-user malleable mobile applications, where end-users can specify requirements that are automatically implemented via LLM-based code generation. However, realizing this vision requires overcoming the key challenge of designing automated test generation that can reliably verify both the presence and correctness of user-specified functionalities. We propose \tool, a user-requirement-driven GUI test generation framework that incrementally navigates the UI, triggers desired functionalities, and constructs LLM-guided oracles to validate correctness. We build a benchmark spanning six popular mobile applications with both correct and faulty user-requested functionalities, demonstrating that \tool effectively validates per-user features and is practical for real-world deployment. Our work highlights the feasibility of shifting mobile app development from a product-manager-driven to an end-user-driven paradigm.

\end{abstract}

%% file: src/ch01-introduction.tex

\section{Introduction}

Software should be malleable~\cite{malleablesystems}, meaning it is easy to change, adaptable, configurable, and modifiable even after shipping to the customers. This notion shares common principles with early adaptive systems~\cite{ashby2013design} from the 1960s, dynamic and reflective programming~\cite{foote1989reflective} during the 1990s, and autonomic computing~\cite{kephart2003vision} in the 2000s. In Kephart and Chess's vision of autonomic computing~\cite{kephart2003vision}, they defined five key aspects of self-management, self-configuration, self-optimization, self-healing and self-protection, which closely align with the principles of malleable software.

Malleable software can be realized in a variety of implementation forms. For example, configurable software systems such as Visual Studio Code and WordPress allow end-users to customize themes and extensions to suit their needs. Similarly, adaptive recommender systems like TikTok and RedNote automatically tailor recommended content based on end-user behavior. In addition, the malleability of GUIs is another widely-studied form of malleable software. GUIs can be malleable either through predefined visual widgets~\cite{gobert2023lorgnette}, flexible placeholders~\cite{chen2025taskartisan} or by generating interfaces on-demand based on end-users' tasks and requirements~\cite{cao2025generative, chen2025taskartisan}. For on-demand GUI generation, the interface adapts to end-user demands. On one hand, end-users may require GUIs that align with specific tasks, where the underlying data sources can vary. Cao et al.~\cite{cao2025generative} leverage LLMs to enable end-users to generate and customize UIs according to their intended tasks and the associated data models using natural language. On the other hand, end-users may have individual preferences for the information and interface elements displayed. Min et al.~\cite{min2025malleable} investigate malleable overview-detail interfaces that allow end-users to tailor views to highlight attributes most relevant to their needs. However, these forms are tailored to specific software and therefore lack wider generalizability.

\begin{figure*}[!t]
    \centering
    \includegraphics[width=1\linewidth]{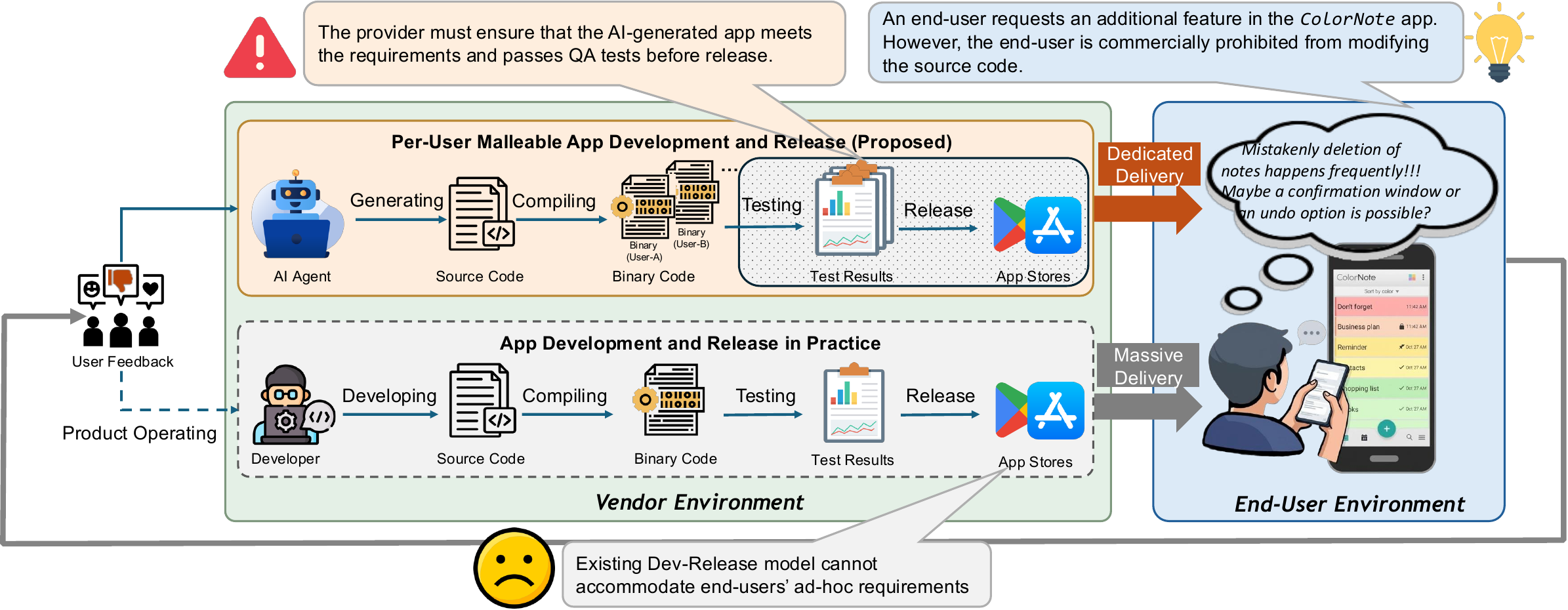}
    \vspace{-10pt}
    \caption{Motivation Scenario and a Conceptual Model for Per-User Malleable Application Development and Release}
    \label{fig:motivation}
\end{figure*}

Differently, we reconsider the problem and opportunities of malleable software within the mobile ecosystem from a broader and more generalizable perspective. Recent decades have seen the success of mobile applications developed in a standardized manner and distributed at scale to broad end-user populations~\cite{applerelease, googlerelease}. 
Mobile applications released to end-users are typically created in a standardized form according to requirements that address the needs of the majority of users and aim to maximize commercial value. However, the unique requirements may only be satisfied when accumulating into significant demands and commercial benefits~\cite{grano2017android, han2012understanding}. Otherwise, they are typically ignored and sacrificed as a trade-off for controlling development costs. In most cases, limited customization is confined to the options provided within the \textit{settings} page. The status quo is reasonable, as it aligns with the traditional software development and release model, in which developers and testing teams are primarily responsible for software implementation and quality. End-users, who generally have limited knowledge of the underlying implementation and limited ability to modify it, interact with the interfaces and functionalities designed and implemented by product managers and developers.

Leveraging recent advances in LLM-based code generation~\cite{zhang2023repocoder, cheng2024dataflow, liu2024graphcoder, wu2024repoformer, zhang2025hierarchical, zhang2024codeagent, wang2025rlcoder, gu2025effectiveness, han2024archcode, mu2024clarifygpt, zhang2025little}, \textit{we envision a future in which mobile application development shifts from being product-manager-driven to end-user-driven, enabling end-users' specific requirements to be implemented quickly and accurately in an exclusively per-user environment, a capability we term a new form of malleable mobile applications}.
Indeed, the emergence of \textit{``vibe coding''}~\cite{ray2025review} has laid the groundwork for a development paradigm where participants are not necessarily professional developers~\cite{vibecoding}. Nevertheless, before this paradigm shift can be fully realized, several critical challenges must be addressed.
Specifically, most mobile applications are proprietary, meaning that end-users do not have access to the source code in their mobile environments. A realistic compromise is for a user to specify his/her requirement to the remote vendor, after which the exact requirement, together with the application's source code, is processed by an LLM in the vendor's environment (see Figure \ref{fig:motivation}). Then, the user-desired functionality is implemented, integrated as a new application version, and delivered back to the end-user. However, despite advances in automatic code generation, \textit{the presence and correctness of user-desired functionality} must be fully validated before release, highlighting the need for an automated testing approach capable of handling high-volume user requests.

In the mobile setting, the testing approach must generate test sequences that \textbf{(a)} efficiently navigate to the entry UI state of the user-desired functionality; \textbf{(b)} verify its presence and trigger it, and \textbf{(c)} validate its correctness after invocation, which falls into the domains of test generation and oracle generation.
Existing LLM-based test generation techniques primarily target unit test generation~\cite{schafer2023empirical, wang2024hits, ryan2024code, cheng2025rug, nan2025test}, which are generally limited to UI-free scenarios.
Meanwhile, many LLM-based oracle generation techniques rely solely on analyzing the software under test (e.g., static syntactic structure or dynamic execution feedback)~\cite{hossain2025togll, hayet2024chatassert, wang2026iterative, zhou2025insights, baral2024automating, johnson2025generating}, rendering them incapable of validating the correctness of user-desired functionalities.
Recently, researchers have explored incorporating Javadoc comments as complementary information to improve oracle generation~\cite{hossain2025doc2oracll}, leveraging LLMs to translate informal natural language prompts into formal fine-grained postcondition assertions~\cite{endres2024can, kande2024security, khandaker2025augmentest}. 
Nevertheless, their oracle construction is centered on constraints over explicit syntactic elements (e.g., variables), instead of high-level semantically-meaningful UI elements.

We propose \tool, a user-requirement-driven GUI test generation framework that addresses the urgent need to validate user-specified functionalities, accelerating this paradigm shift. \tool operates in three phases: (1) \textbf{Functional Triggering State Navigation} incrementally explores the UI state space to identify an entry state and interaction path that are semantically aligned with the given requirement, guided by an LLM-based relevance scoring mechanism; (2) \textbf{Functional Presence Check \& Execution} generates and executes triggering scripts to verify the existence of the desired functionality and ensure it can be successfully invoked through valid UI interactions, meanwhile producing intermediate assertions; (3) \textbf{Functional Correctness Oracle Generation} constructs test oracles by comparing pre- and post-execution states, capturing behavioral differences to determine whether the invoked functionality correctly satisfies the requirement. \tool further incorporates a selector refinement mechanism across all phases to detect and repair invalid or ambiguous UI element selectors.

We construct a dataset spanning six popular mobile applications and 34 user requirements collected from Google Play. Based on this dataset, we further build a benchmark comprising 144 application versions that correctly implement 34 user requirements across the six apps, as well as 64 compilable versions injected with faulty functionalities. We evaluate \tool on the correct versions, achieving an accuracy of 81.2\%. We further assess \tool on a mixed dataset containing both correct and faulty versions, where it attains an average precision and specificity of 90.2\% and 89.1\%, respectively. In addition, we conduct a comparative study to demonstrate its superiority over existing approaches, and perform a performance evaluation to assess its practicality in real-world scenarios. \todo{We also evaluate \tool's overhead and sensitivity.}

Our paper makes the following contributions.

\begin{itemize}[leftmargin=*]
    \item We demonstrate the practical feasibility and promote per-user malleable mobile applications, enabling end-users' requirement-driven functionality automatically implemented.
    \item We propose \tool, a requirement-driven GUI test generation framework for validating the correctness of user-desired functionality implementation.
    
    \item We construct a benchmark dataset of real mobile apps with correct and faulty user-requested functionalities and demonstrate \tool's effectiveness and practicality.
\end{itemize}

%% file: src/ch02-motivation.tex
\section{Malleable Software in the LLM Era}

Traditionally, software is largely predefined at design time, following a lifecycle of \textit{requirements}-\textit{implementation}-\textit{release}. Once deployed, modifying functionality typically requires re-entering a costly and time-consuming development cycle involving multiple vendor-side stakeholders, including product operators, developers, and testers. Consequently, software behavior is predominantly static and provider-controlled. \textit{Malleable software}~\cite{malleablesystems} shifts this paradigm by relaxing the assumption that system behavior must be fully determined at design time. Instead, functionality can be adapted or extended after deployment, enabling systems to better accommodate individual user requirements. Both developers and end-users can participate in shaping system behavior on demand. Since user experience~\cite{al2023people} is the foundation of software, malleable software can significantly enhance this experience, thereby contributing to the success of the system and potentially reshaping the entire ecosystem. We identify key characteristics that enable malleable systems while preserving mutual benefits for providers and end-users: \textbf{\textbf{End-user-driven Customization:}} End-users can influence and reshape system behavior based on their specific needs; \textbf{\textbf{Natural Language Interface:}} Natural language acts as a core interface for specifying, modifying, and composing functionality; \textbf{\textbf{Source Code Access.}} Functional modifications should be achieved without direct access of the underlying source code; \textbf{\textbf{Reliability:}} The customized system should maintain the same level of reliability as the original system after modification; \textbf{\textbf{Vendor Privilege:}} The vendor owns the intellectual property~and retains full authority to audit and approve customized~versions.

The unleashed potential of LLMs has created new opportunities for implementing malleable systems. The recent advancements of ``vibe coding''~\cite{ray2025review} highlights a possible approach in which the code generation process can be fully automated by providing a natural language description of requirements. We envision a future in which mobile applications become easily malleable, enabling them to satisfy the specific needs of individual end-users. Recent literature has conducted preliminary explorations of malleable systems, with a primary focus on user interfaces. Notable examples include task-driven UIs with generative form widgets~\cite{cao2025generative}, which respond to user-specified natural language task descriptions; the infinite canvas~\cite{chen2025taskartisan}, where users can place interactive widgets (e.g., images, notes) using sketches and voice to support sensemaking tasks; and user-need-oriented overview-detail interfaces~\cite{min2025malleable}, which allow end-users to selectively show or hide specific information according to their needs. These examples represent efforts to make user interfaces malleable, adaptable, and responsive to individual users. However, these systems remain rigid, as they are merely designed around the UIs instead of the back-end implementations, and designed within specific frameworks, scenarios, and constraints.

\textbf{Motivating Scenario.} Figure \ref{fig:motivation} illustrates a motivating scenario and a conceptual model for per-user malleable application development and release. Typically, an end-user may request specific features in an existing app. For example, consider a real-world complaint from a user of \textit{ColorNote}, who reported that ``\textit{notes are frequently deleted by mistake}'' and suggested that ``\textit{a confirmation window or an undo option would be helpful}.'' Typically, the vendor does not provide access to its proprietary code and is commercially prohibited from allowing users to modify the source code directly. Under the conventional development-release model, user feedback is collected, and the requested feature may be implemented only if the vendor prioritizes it over other requests. The app is then massively re-distributed as a unified binary. It has several drawbacks: \textit{(i)} it neglects the long tail of user demands, leaving a substantial portion of requests unsatisfied; \textit{(ii)} the remaining end-users must bear the cost of additional functionality, including increased cognitive load (e.g., more settings options) and higher resource consumption, although individually they do not require the particular feature. 

To highlight users' varied requests, we select ten representative user comments that require specific feature enhancements on the original app, presented in Table~\ref{tab:features}. The user comments are lightly paraphrased for better readability while preserving the original intent. These requests span a diverse set of application domains and primarily focus on feature augmentation (e.g., search, sharing), interaction improvements (e.g., confirmation and undo mechanisms), personalization options (e.g., language support, displaying),~etc.

In contrast, we envision a per-user malleable app development and release model that can satisfy individual user demands while preserving the vendor's control over proprietary code. The proposed framework addresses this gap by enabling per-user customized app generation. First, user feedback is promptly collected and returned to the vendor's environment. Next, a LLM agent generates customized source code for each individual user and compiles it into a dedicated binary. The vendor then tests the customized app before releasing it to the corresponding user. However, a tension arises between full automation in the face of a high volume of user requests and ensuring the quality of the customized app, such as whether the functional requirements are fully satisfied or whether non-functional properties are compromised.

\input{src/sub/features}

\textbf{Open Challenges.} We raise several open challenges in addressing per-user malleable app development and release:

\begin{itemize}[leftmargin=*]
    \item \textbf{End-user-driven Code Generation:} Generating reliable and correct code directly from end-user requirements expressed in natural language. Challenges include understanding ambiguous requests, mapping user intent to functional features, and producing code that integrates correctly with existing systems.
    
    \item \textbf{Functional Testing:} Ensuring that the generated app behaves correctly according to the specified functional requirements. This requires generating test inputs and oracles that accurately capture the intended behavior of user-requested features.
    
    \item \textbf{Non-functional Testing:} Verifying non-functional properties such as performance, security, and usability.
    
    \item \textbf{Regression Testing:} Ensuring that new per-user customizations do not break existing functionality. Efficiently performing regression tests for potentially large numbers of individualized app instances is a significant challenge.
\end{itemize}

%% file: src/sub/features.tex
\begin{table}[!t]
\centering
\footnotesize
\caption{Frequently-mentioned User Requirements}
\label{tab:features}
\vspace{-10pt}
\renewcommand{\arraystretch}{1.2}
\begin{tabular}{p{0.5cm}p{2.5cm}p{2cm}p{8.5cm}}
\toprule
\textbf{ID} & \textbf{App Name} & \textbf{Category} & \textbf{User Comment} \\
\midrule
1 & Acrobat Reader & Productivity & I wish the app supported more language options, especially Bengali. \\
\hline
2 & PDF Viewer & Productivity & I want a search function to search within the current PDF page. \\
\hline
3 & ColorNote & Productivity & I want a confirmation window before deleting a note to avoid mistouch. \\

\hline
4 & Blogger & Social & I wish the app shows the correct publish time when I publish a post. \\
\hline
5 & Groovebook & Photography & I want the app to provide a picture resizing option. \\
\hline
6 & Golf GPS & Sports & I want a share button. \\

\hline
7 & Facebook & Social & I want the feed page to offer sorting options based on posting time. \\
\hline

8 & Google News & News & I wish the app had a dark mode option. \\
\hline
9 & Google News & News & I want the app to support both automatic refresh and manual refresh. \\

\hline
10 & Amazon Kindle & Books & I wish the app provided options to adjust paragraph spacing for better readability. \\
\bottomrule
\end{tabular}

\end{table}

%% file: src/ch03-approach.tex
\section{Approach}\label{sec:approach}

We address the challenge of \textbf{functional testing} by proposing \tool, composed of three phases.

\begin{figure*}[!t]
    \centering
    \includegraphics[width=1\linewidth]{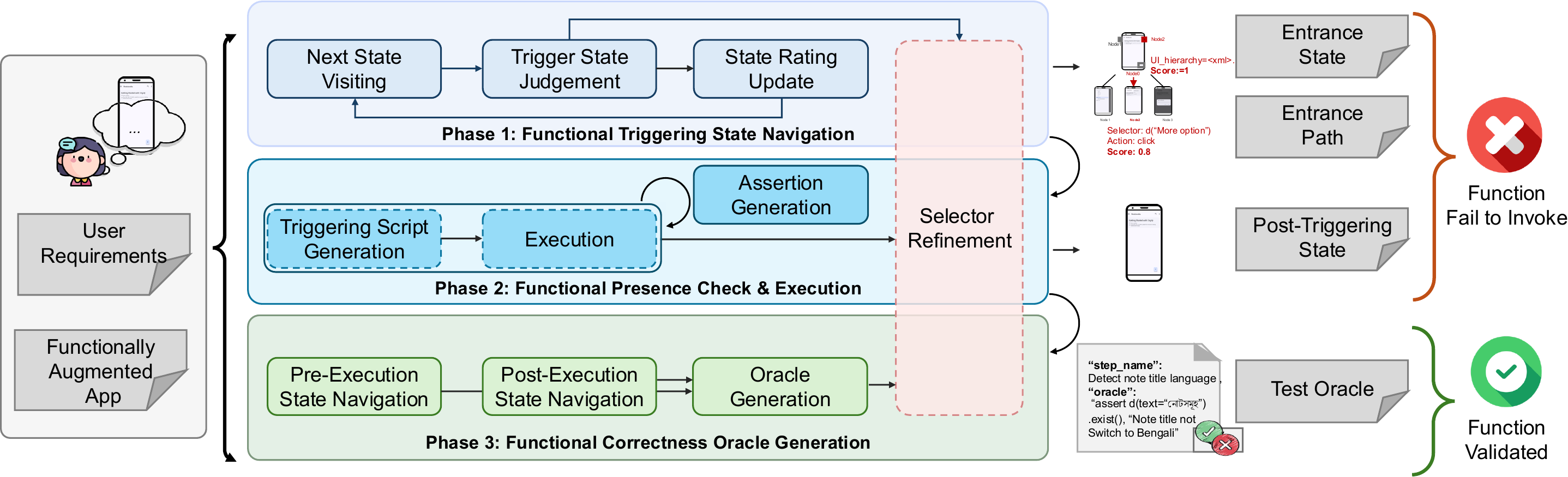}
    \vspace{-10pt}
    \caption{Overview of \tool}
    \label{fig:framework}
\end{figure*}

\subsection{Problem Definition}

We address the challenge of \textbf{functional testing} for validating the functional correctness of \emph{malleable software}, where end-users specify desired requirements in natural language, and an automated agent (e.g., LLM-based agent) augments the original application with the additional functionality. We focus our analysis on GUI-based applications, which represent the most common form of application for end-users. Given a natural language requirement $r$ and a functionally-augmented application $\mathbb{A}$, our objective is to automatically \textbf{validate whether $r$ is correctly implemented in $\mathbb{A}$}. To assess the correctness of the newly implemented functionality, our approach proceeds in three steps.

     \noindent \textbf{Navigation:} Identify and reach the UI state containing the element that can trigger the functionality corresponding to $r$.
    \noindent \textbf{Verifying and Triggering:} Using the identified UI state and element as input, verify that the corresponding option is mapped to $r$, perform the appropriate action to activate the functionality.
    \noindent \textbf{Observation and Validation:} Monitor the application's response upon triggering the action, generate the oracle that can confirm the correctness of the user-desired functionality from~$r$.

To validate the \textbf{functional testing} of $\mathbb{A}$, we expect the test case consisting of three fundamental components, defined as a tuple: $T = (P, E, \mathcal{O})$, where $P$ denotes the \emph{precondition state operations}, i.e., a sequence of operations (actions on the corresponding UI elements) that navigates the application to the entry state of triggering the new functionality, $E$ represents the \emph{execution state operations}, i.e., a sequence of operations that check and trigger the corresponding functionality (e.g., click after text inputs), and $\mathcal{O}$ corresponds to the \emph{post-execution oracle}, which determines whether the application realizes the user-desired functionality correctly after the functionality has successfully executed.

\subsection{Framework Overview}

We propose \tool, a requirement-driven framework for generating GUI test cases to validate the functional correctness of applications in scenarios where they are automatically augmented with user-desired functionalities. Figure~\ref{fig:framework} presents an overview of our approach.
Our framework consists of three key components:

\begin{itemize}[leftmargin=*]
    \item \textbf{Functional Triggering State Navigation:} This phase incrementally navigates 
    on the GUI states, prioritizing and identifying states that are likely to reach new-functionality-trigger entrances relevant to the given requirement.

    \item \textbf{Functional Presence Check \& Execution:} Given the identified entry state, this phase iteratively generates and executes triggering scripts to reach the target functionality, while deriving assertions to ensure that the resulting UI elements are consistent with the requirement.

    \item \textbf{Functional Correctness Oracle Generation:} This phase constructs test oracles by obtaining pre- and post- execution states, and comparing their differences to validate whether the triggered functionality correctly satisfies the requirement.
\end{itemize}

Phase 2 and Phase 3 separate the tasks of functional presence check (whether the user-desired functionality is present) and functional correctness validation (whether it has been implemented correctly). Furthermore, selector refinement is applied in each phase to detect and correct wrong UI elements, properties, and actions.

\subsection{Functional Triggering State Navigation}\label{sec:app:1}

Given a requirement $r$ and a functionally augmented application $\mathbb{A}$, \tool aims to navigate the application from an initial UI state (e.g., the home page) to an entry page where it can trigger the user-desired functionality corresponding to $r$. Each UI state $s$ is associated with a set of UI operations $Op$. An operation $op \in Op$ is defined as a 2-tuple $\langle ele, act \rangle$, where $ele$ represents a UI element and $act$ represents an interaction action applicable to that element (e.g., click, swipe, long-click). Through dynamic navigation, \tool efficiently explores the UI state space while evaluating candidate interactions guided by their semantic relevance to the requirement, reaches the UI state that can trigger the user-desired functionality, and identifies the corresponding operation. Notably, to reduce both the extensive state space and the computational cost associated with random or fixed exploration methods (e.g., BFS), we adopt a heuristic-driven scoring mechanism that directs the exploration toward the target state more effectively. The overall process is presented in Function \ref{alg:phase1}. Upon completion, \tool returns the entry state $s^{\prime}$, the exploration history $H$, and the corresponding triggering operation $Op$ for the user-desired functionality.

\input{src/sub/alg1}

\textbf{Next State Visiting.} Function lines 7 to 15 present the process of visiting next state. Function \ref{alg:phase1} lines 7-15 describe the procedure for visiting the next state during exploration. Specifically, it begins by selecting a to-be-visited entry $n$ from the priority queue. The entry consists of a base state $n.s$ where the operation $n.op$ will be performed on, and a relevance score $n.\Gamma$. If the base state $n.s$ differs from the current state, it switches the current state to $n.s$ by replaying the execution path from the initial state $s_0$ to $n.s$ using \texttt{Replay}. Subsequently, it checks whether the next state resulting from applying operation $n.op$ to the base state $n.s$ has already been explored using \texttt{EquivalentState}. If so, the to-be-visited entry is skipped to avoid unnecessary exploration. Otherwise, \tool performs the operation $n.op$ on state $n.s$ to reach a next state $s^{\prime}$. Meanwhile, the state-operation pair $\langle n.s, n.op \rangle$ is recorded in the history $H$. Finally, the current state is updated to $s^{\prime}$.

\texttt{EquivalentState}. Given the history $H$, the base state $n.s$, and the operation $n.op$, we determine whether the next state resulting from applying $n.op$ to $n.s$ has already been visited in the history. Since the same state may be reached via different paths, we use the combination of the base state $n.s$ and the operation $n.op$ as a unique identifier. Formally, the EquivalentState function computes the equivalence result (es), defined as:

\begin{equation}
\text{es} =
\begin{cases}
1, & \text{if } \exists s_x \in H \ \text{s.t.} \ \text{equal}(s_x, n.s) \\
   & \quad \text{and } \exists op_x \in \text{edge}(s_x, H) \ \text{s.t.} \ \text{equal}(op_x, n.op), \\
0, & \text{otherwise.}
\end{cases}
\end{equation}

Here, $\text{edge}()$ returns the set of operations connected to a given state in $H$, and $\text{equal}()$ determines whether a state or operation matches another by comparing their hash values.

\textbf{Trigger State Judgement.} Function lines 16 to 19 describe the trigger state judgment process. 
Given the reached next state $s^{\prime}$, \tool invokes \texttt{PageExplorer} with the requirement $r$ to determine whether $s^{\prime}$ is an entry state capable of triggering the user-desired functionality. The procedure returns a boolean flag $isEntry$ along with a set of candidate operations $Op$. If $isEntry$ is true, \tool terminates and returns the corresponding entry state $s^{\prime}$ for triggering the user-desired functionality, along with the exploration history $H$, and the triggering operations $Op$. Otherwise, the exploration process proceeds. 

\begin{figure}[!t]
    \centering
    \includegraphics[width=\linewidth]{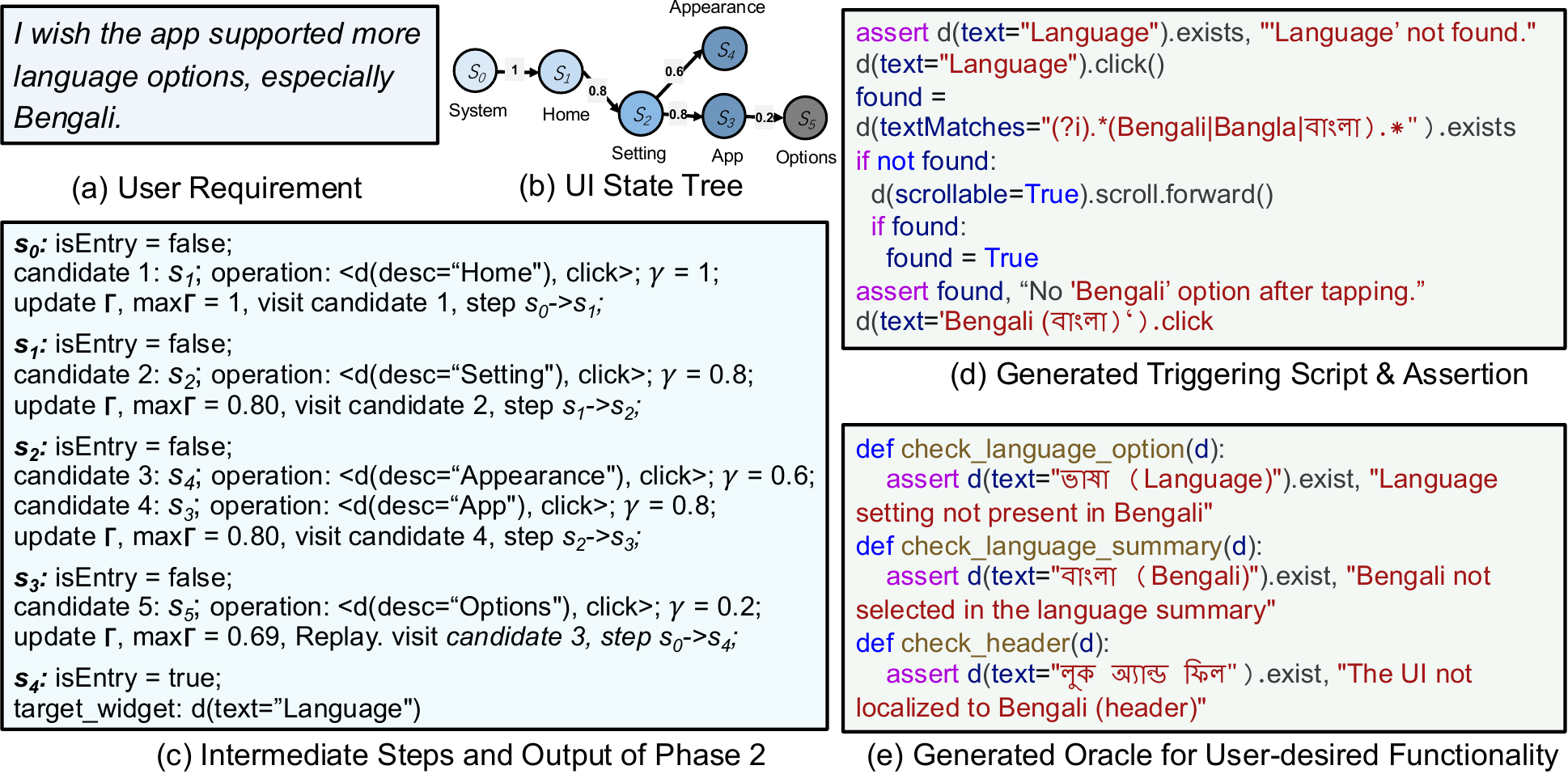}
    \vspace{-11pt}
    \caption{Illustrative Example}
    \label{fig:example}
\end{figure}

\begin{figure}[!t]
    \centering
    \includegraphics[width=0.85\linewidth]{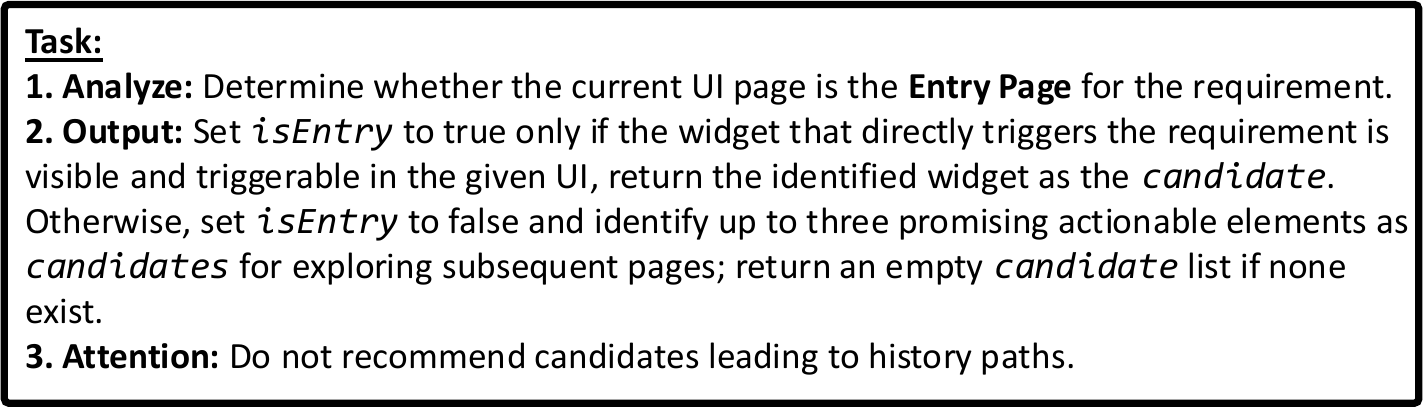}
    \vspace{-10pt}
    \caption{Prompt for Trigger State Judgement}
    \label{fig:prompt}
\end{figure}

\texttt{PageExplorer}. Given a requirement $r$ and a newly reached state $s^{\prime}$, we leverage an LLM to determine whether there exists a UI element $ele$ and a corresponding action $act$ that can trigger the user-desired functionality. To facilitate this process, we construct a prompt based on the UI DOM tree of $s^{\prime}$. To reduce redundancy and avoid attention dispersion, we compress the DOM tree by retaining only semantically meaningful attributes (e.g., \textit{text}, \textit{content-desc}, and \textit{resource-id}) while pruning irrelevant information. The simplified prompt (excluding input and output format specifications) is presented in Fig.~\ref{fig:prompt}. If $isEntry$ is false, the LLM is expected to return a set of $k$ candidate operations.

\textbf{State Rating Update.} Function lines 20 to 23 describe the process of updating state ratings. For each candidate operation $op \in Op$ returned by \texttt{PageExplorer}, \tool computes a relevance score $\Gamma$ via \texttt{computeScore}. Each resulting tuple $\langle s^{\prime}, op, score \rangle$ is then inserted into the priority queue, prioritizing more promising state-operation pairs in subsequent steps.

\texttt{computeScore}. Given a requirement $r$, a base state $s^{\prime}$, an operation $op$, and the history $H$, we leverage the semantic capability of an LLM to compute the relevance score $\Gamma$. The relevance score of operation $op$ in state $s^{\prime}$ is calculated as the cumulative path score from the initial state $s_0$ to $s^{\prime}$ using the geometric mean:

\begin{equation}
    \Gamma_{s^{\prime}, op} = \left( \prod_{i=1}^{d} \gamma_i \right)^{\frac{1}{d}}
\end{equation}

where $\gamma_i$ denotes the atomic relevance score of state $s_i$ under operation $op_i$, and $d$ represents the depth of the path from $s_0$ to $s^{\prime}$. This approach reflects the sequential nature of GUI navigation, where using the geometric mean reduces the impact of low-confidence steps, ensuring that low-score elements are not overshadowed by high-score ones compared to arithmetic averaging. Each $\gamma_i \in [0.2, 1]$ is a discrete, five-level value representing the semantic likelihood that executing operation $op_i$ from state $s_i$ will lead toward the user-desired functionality. The detailed scoring rules are presented in Table~\ref{tab:score-rule}. We normalize the relevance scores to facilitate computation. These rules serve as guidelines for the LLM to identify the most related score category. 

\textbf{\textit{Example.}} Figure~\ref{fig:example}(a) illustrates a user requirement: adding language support for Bengali. Given an implemented version of the application guided by the user requirement, \tool first navigates toward the functionality-triggering state, with the UI state transitions shown in Figure~\ref{fig:example}(b). Starting from the system page, \tool enters the home page of the app \texttt{Orgzly\_revived} (notes app), clicks the \textit{Setting} button, then the \textit{App} option, go backwards, and proceeds by selecting the \textit{Appearance} option. Figure~\ref{fig:example}(c) presents the intermediate steps and corresponding outputs. At state $s_3$, \tool observes that entering candidate state 3 has a higher relevance score than candidate 5, indicating a more promising exploration path. Therefore, \tool goes back to state $s_2$ and proceeds to visit $s_4$. This example demonstrates how relevance scores guide \tool to effectively navigate toward relevant UI states.

\input{src/sub/score_rule}

\subsection{Functional Presence Check \& Execution}\label{sec:app:2}

Given the identified entry state $s^{\prime}$ for invoking the user-desired functionality, this phase verifies whether the functionality contains the necessary components and can be triggered through a sequence of interaction actions. Specifically, \tool generates and executes a sequence of actions to interact with relevant UI elements (e.g., buttons and text fields), progressing according to the intended functionality. This process may be iterative, as some functionalities require multiple interaction steps rather than a single action. Meanwhile, \tool performs assertion generation to validate that the resulting state contains the required UI elements and actions corresponding to the requirement $r$.

\textbf{Triggering Script Generation \& Execution.} This step generates a triggering script based on the current state $s'$ and the associated operation $op$. The script is then executed to transition to the next state $s''$, followed by assertion generation to verify the presence of required UI elements. If necessary, the process iteratively continues with further script generation to advance toward the user-desired functionality. The process terminates when the resulting state $s''$ satisfies the completion condition of the target functionality under requirement $r$. The state transition from $s'$ to $s''$ can be represented as $s' \rightarrow \dots \rightarrow s''$, with the corresponding operations denoted by $Op'$.

\textbf{Assertion Generation.} Given the newly reached state $s''$, this step generates assertions to verify that the UI elements required by the user-desired functionality are present in $s''$. For example, asserting that the ``Bangladesh'' language option is present in the list view element (e.g., \textit{\texttt{assert} d(textMatches="(?i).*(Bengali|Bangla).*").exists, "the language selector did not show an option matching 'Bengali'"}).

The processes are implemented by guiding the LLM with a structured, role-based prompt that enforces detailed API usage rules (e.g., \texttt{uiautomator2} compliance), explicit action-assertion patterns, and anti-hallucination measures, ensuring that the generated outputs are both reliable and executable. Upon completion of the above steps, $\mathbb{A}$ completes user-desired functionality invocation and transitions to the post-triggering state.

\textbf{\textit{Example.}} Figure~\ref{fig:example}(d) presents the generated triggering script and corresponding assertions derived from the user requirement and the target application. The script first asserts the existence of the \textit{Language} button and then performs a click action. Ideally, this action brings up the language selection page. Subsequently, the script verifies whether the selection list contains entries such as \textit{Bengali} or \textit{Bangla}. It then continues to scroll through the list until a matching entry is found. Finally, a click action is invoked to select the desired language, triggering the user-desired functionality.

\subsection{Functional Correctness Oracle Generation}

This phase determines whether the requirement $r$ has been correctly implemented.

\textbf{Pre-Execution and Post-Execution State Navigation.} To assess whether the user-desired functionality is successfully triggered, we need to compare state both before and after the user-desired functionality is triggered. For example, consider requirement 3 in Table~\ref{tab:features}, where the user requests a confirmation window; Phase 2 (Section~\ref{sec:app:2}) has already validated the existence of the window; In this phase, we need to compare the pre-execution and post-execution states after the confirmation option (Yes or No) is selected. The process begins from the initial state $s_0$ and navigates to the matched state in a manner similar to Phase 1 (Section~\ref{sec:app:1}). Upon completion, we obtain the pre-execution state $s_{pre}$ and the post-execution state $s_{post}$.

\textbf{Oracle Generation.} Given the pre-execution state $s_{pre}$, the post-execution state $s_{post}$, and the executed operations $Op'$ as functionality records, \tool generates an oracle to validate whether the user-desired functionality has been correctly implemented. Since the functionality may be partially or imperfectly implemented, multiple sub-oracles are generated to fully capture correctness. Let $\eta$ denote the pre-defined number of sub-oracles. The final oracle set is then represented as:

\begin{equation}
    \mathcal{O} = \{ o_1, o_2, \dots, o_\eta \}
\end{equation}

\textbf{\textit{Example.}} Figure~\ref{fig:example}(e) presents the generated oracle for the requirement in Figure~\ref{fig:example}(a). The three assertions verify the Bengali language option, the currently selected language, and the Bangladesh string in the word ``settings''.

\subsection{Selector Refinement}

To mitigate selector hallucinations or ambiguities introduced by the LLM in UI element selection, we integrate a UI element selector refinement process that automatically detects and repairs failed selectors. Specifically, \tool first extracts key attributes such as \texttt{resource}\texttt{-id}, \texttt{text}, and \texttt{content-desc} from failed selector. Then, \tool performs hierarchical search to match the exact attribute that exists in the current UI state DOM, following a structured order from exact attribute name match, cross-attribute name match (e.g., matching \texttt{text} with \texttt{content-desc}), fuzzy name inclusion (partial value matches), to class name match. Finally, a more accurate and stable selector is generated, prioritizing elements including \texttt{resource-id}, \texttt{text}, and \texttt{content-desc}.


%% file: src/sub/alg1.tex
\begin{algorithm}[!t]
\caption{Requirement-Guided Dynamic Navigation}
\label{alg:phase1}
\begin{algorithmic}[1]
\STATE \textbf{Input:} requirement $r$, initial state $s_0$, max steps $M$
\STATE \textbf{Output:} entry state $s_{\prime}$, exploration history $H$,

\STATE $V \gets \emptyset$, $H \gets \emptyset$, $curr\_state \gets \emptyset$
\STATE PushPriQueue($\langle s_0, \langle btn, \texttt{click} \rangle,1 \rangle$)

\WHILE{$Q \neq \emptyset$ and step $<M$}
    \STATE $n \leftarrow$ PopPriQueue()
    \STATE $step \leftarrow step +1$
    \IF{$n.s \neq curr\_state$}
        \STATE Replay($s_0$, $n.s$)
    \ENDIF
    \IF{EquivalentState$(n.s, n.op, H)$}
        \STATE continue
    \ENDIF
    \STATE Append($H$, $n.s$, $n.op$)
    \STATE $s^{\prime} \leftarrow$ Visit($n.s$, $n.op$) 
    \STATE $curr\_state \leftarrow s^{\prime}$
    \STATE $isEntry, Ops \leftarrow$ PageExplorer($r$, $s^{\prime}$)
    \IF{$isEntry$}
        \STATE return $(s^{\prime}, H, Ops)$
    \ENDIF
    \FOR{each candidate $op \in Ops$}
        \STATE score $\leftarrow$ computeScore($r$, $s^{\prime}$, $op$, $H$)
        \STATE PushPriQueue($\langle s^{\prime}, op, score \rangle$)
    \ENDFOR
\ENDWHILE
\STATE return $(null, H, null)$
\end{algorithmic}
\end{algorithm}

%% file: src/sub/score_rule.tex
\begin{table}[!t]
\centering
\footnotesize
\caption{Relevance Scoring of Candidate UI Elements}
\vspace{-10pt}
\label{tab:score-rule}
\renewcommand{\arraystretch}{1.2}
\begin{tabular}{m{2.4cm}m{2.2cm}m{8cm}m{0.5cm}}
    \toprule
    Requirement Relevance & Criteria & Description & $\gamma*5$ \\
    \midrule
    \shortstack[l]{Highly likely} & \shortstack[l]{Literal equivalence} & The text of the UI element $op.ele$ provides a clear and unambiguous mapping to the requirement text. & 5 \\
    \shortstack[l]{Strongly related} & \shortstack[l]{Semantic equivalence} & The text of the UI element $op.ele$ has a clear semantic relation to the target functionality  & 4 \\
    \shortstack[l]{Possibly related} & \shortstack[l]{Structural relation} & The element $op.ele$ is a parent container logically related to the target functionality. & 3 \\
    \shortstack[l]{Weakly related} & \shortstack[l]{Generic aggregation} & The UI element $op.ele$ links to a generic aggregation page that may contain a large number of unexplored pages. & 2 \\
    \shortstack[l]{Irrelevant} & \shortstack[l]{No relation} & The UI element has little or no relation to the user-desired functionality. & 1 \\
    \bottomrule
\end{tabular}
\end{table}

%% file: src/ch04-evaluation-setup.tex
\section{Evaluation Setup}

We design the following research questions:

\begin{itemize}[leftmargin=*]
    \item \textbf{Success Rate (RQ1)}: How is the success rate of \tool?
    \item \textbf{Discrimination Capability (RQ2)}: Can \tool generate oracles that distinguish correct implementations from faulty ones?
    \item \textbf{Comparison (RQ3)}: How does \tool compare to the corresponding stage-wise baselines?
    \item \textbf{Overhead (RQ4)}: How is the overhead of~\tool?
    \item \textbf{Parameter Sensitivity (RQ5)}: How do parameter affect \tool's effectiveness?
\end{itemize}

\textbf{Data Construction.} Each entry in our dataset is represented as a tuple $\langle r, \mathbb{A}, correct \rangle$, where $r$ denotes the requirement, $\mathbb{A}$ denotes the functionally augmented application, and $correct \in \{\textit{true}, \textit{false}\}$ indicates whether $\mathbb{A}$ correctly implements the functionality specified by $r$. A \textit{correct} implementation satisfies the target requirement, whereas a \textit{faulty} implementation fails to satisfy the requirement, although it can be compiled, installed, and executed normally. The construction process comprises three steps.

(a) \textbf{Requirement collection.} We collect candidate requirements from the public Google Play review dataset\cite{lava18googleplay}. Specifically, we randomly sample 1,000 user comments. We adopt a two-stage filtering process. First, we leverage the semantic understanding capability of an LLM to remove purely subjective praise, generic complaints, and comments without functional intent, which yields 533 candidate comments. Subsequently, we manually inspect the remaining comments and remove those that are too vague to constitute a concrete functional expectation, request bug fixes, or cannot be translated into an actionable requirement. We retain only requirements that can be expressed as executable functionality, resulting in 34 natural-language requirements. The labeling process was conducted by two of the authors over two iterations, achieving an overall Cohen's Kappa of 0.72. Table~\ref{tab:req-taxonomy} presents the overall statistics of our collected requirements. \textit{General} requirements refer to reusable functionalities like theme, browsing, file handling, language, notifications, account, and sharing and \textit{Domain-Specific} requirements refer to functionalities with stronger domain semantics, such as health management or content creation.

\input{src/sub/table_requirements}

(b) \textbf{Mapping Requirements to Selected Apps.} We select six representative open-source Android apps from F-Droid, covering the domains of the collected requirements: \textit{Einkbro} (browser), \textit{Foodyou} (health), \textit{Fossify\_camera} (camera), \textit{News\_Reader} (news), \textit{Orgzly\_revived} (notes), and \textit{Tusky} (social). We then map each requirement to the appropriate apps, yielding 204 requirement-app pairs. If the original app version already satisfies a requirement, we directly create a dataset entry and label it as a \textit{correct} implementation. If the functionality can be implemented but is currently absent, we label the corresponding pairs as \textit{unimplemented}. Otherwise, if the functionality is not feasible within the app, we label the pairs as \textit{impossible to implement} and exclude them from further analysis. As a result, we obtain \numcorrect correctly implemented requirement-app pairs and \numfaulty unimplemented requirement-app pairs.

\todo{(c) \textbf{Creating Implementation Versions.} For the 80 unimplemented requirement-app pairs, we employ code agents (i.e., Doubao-1.6) as implementation assistants, followed by manual completion and validation of the modifications. 
To mitigate potential issues in agent-generated code, such as logical inconsistencies, stylistic divergence, or incomplete functionality, all generated implementations were carefully reviewed and revised by the authors to ensure consistency with the original design patterns and coding conventions of the target applications. 
During the construction process, we systematically collected and recorded intermediate implementations, including those exhibiting iterative defects or incorrect behaviors. These versions were then manually analyzed and categorized based on their error types. From this pool, we selected implementations that remained compilable but failed to satisfy the intended requirements, and further refined them into representative faulty versions to better reflect realistic error scenarios encountered in practice. }

\todo{The overall construction pipeline involves agent-assisted generation, manual completion and refinement, iterative validation, and faulty version curation. The process was carried out by three authors over a period of more than two months, with over 400 LLM invocations, resulting in a high-quality dataset that supports reliable experimental evaluation.}

As a result, we obtain \numcorrect \textit{correct} implementations and \numfaulty \textit{faulty} implementations. The \textit{faulty} implementations arise from two sources: (1) \textit{Intermediate faulty versions}, i.e., intermediate versions generated during the implementation process that remain executable but do not yet satisfy the requirement; and (2) \textit{Unsuitable attempted implementations}, i.e., versions produced from technically feasible implementation attempts whose resulting behavior does not constitute a valid realization of the target requirement.

Table~\ref{tab:benchmark} summarizes the distribution of dataset cases, including \textit{correct} (C) and \textit{faulty} (F) implementations, across applications and requirement categories. The applications are labeled A-F in alphabetical order.

\input{src/sub/req_detail}

\textbf{Baselines.} We adopt \emph{stage-wise baselines} to isolate the contribution of each phase, as the three phases address different sub-problems and should be compared under their respective objectives. For Phase 1, we use a \emph{prompt-only navigation} baseline (\navbase). Unlike our method, it removes the scoring-based next-state navigation and predicts the next navigation action solely from the current GUI state. For Phase~2 and Phase~3, we adopt \oraclebase~\cite{khandaker2025augmentest}, a representative LLM-based oracle generation approach that uses context-aware prompting. For fair comparison, we preserve the same model and execution environment, making only minimal adaptations for our setting, including replacing the original Java unit-test context with Android GUI context, modifying prompts to adopt Python-based \texttt{uiautomator2}-style assertions.

\textbf{Stage-wise Comparison Protocol.} To answer \textbf{RQ3}, we evaluate each phase independently using the corresponding stage-wise baseline. This design follows the structure of \tool and enables us to measure the effectiveness of each phase without being affected by errors propagated from later phases.

For \textbf{Phase~1}, we compare \tool with \textsc{Prompt.base} on all \textit{correct} cases (\numcorrect in total), as Phase~1 evaluates navigation toward a valid entrance state, which can be sufficiently assessed on correct implementations. For \textbf{Phase~2} and \textbf{Phase~3}, we compare \tool with \oraclebase on both correct and faulty cases, since these phases evaluate whether the generated oracles can correctly validate functional triggerability and post-interaction effects. 

For \textit{correct} cases, the evaluation size of each phase equals the total number of correct cases in the benchmark. For \textit{faulty} cases, we categorize them by the phase at which the implementation failure manifests. Among the 64 faulty cases, 6 lack a valid trigger state, 33 can reach the trigger state but fail to activate the intended functionality, and 25 successfully activate the functionality but produce incorrect results. Accordingly, the latter two groups are used as the faulty evaluation sets for Phase~2 and Phase~3, respectively. During evaluation, all preceding phases are executed correctly, thus any failure is attributable within the local phase.

\textbf{Metrics.} To answer \textbf{RQ1}, we report the success rate (\textit{Succ.}) for each phase and end-to-end. Specifically, \textit{Phase~1} success means reaching the entrance state that can trigger the target functionality; \textit{Phase~2} success means generating a trigger script and assertion that correctly validate its triggerability; \textit{Phase~3} success means generating an oracle that correctly validates its effect after interaction; and end-to-end success means succeeding in all three phases and producing a correct final oracle decision. 

To explore \textbf{RQ2}, we adopt the following metrics:  $\mathit{Precision} = \frac{TP}{TP+FP}, \mathit{Recall} = \frac{TP}{TP+FN}, \mathit{Specificity} = \frac{TN}{TN+FP}$.
Here, a true positive (TP) occurs when a correct version is judged as \textit{pass}; a true negative (TN) occurs when a faulty version is judged as \textit{fail}; a false positive (FP) occurs when a faulty version is judged as \textit{pass}; and a false negative (FN) occurs when a correct version is judged as \textit{fail}.

For \textbf{RQ3}, we use metrics that match the objective of the compared phase. \todo{For \textbf{Phase~1}, we use the same success rate as in RQ1 and additionally report the \textit{distribution of successful steps}, i.e., the number of interaction steps needed to reach the target entrance state over successful runs only, to compare the efficiency of \navbase and \tool.} For \textbf{Phase~2} and \textbf{Phase~3}, since the comparison concerns the quality of generated assertions/oracles on both correct and faulty implementations, we use the same \textit{Prec.}, \textit{Rec.}, and \textit{Spec.} metrics as in RQ2.

\textbf{Environment.} Each benchmark case is associated with an installable APK. Both GUI interaction and oracle validation are performed through \texttt{uiautomator2}\cite{uiautomator2}. 
We use the same environment for fair comparison. We use \texttt{GPT-5} as the state-of-the-art LLM with a temperature of 0. We provide the complete prompt in \cite{anonymous}. The experiments are conducted on a \textit{Pixel\_7} emulator running Android~16. The maximum exploration step $M$ is \todo{5}.

%% file: src/sub/table_requirements.tex
\begin{table}[!t]
\centering
\footnotesize
\caption{Statistics of our Collected Requirements}
\vspace{-10pt}
\label{tab:req-taxonomy}
\setlength{\tabcolsep}{4pt}
\begin{tabular}{p{2.0cm}p{2cm}p{5.0cm}c}
\toprule
\textbf{Group} & \textbf{Category} & \textbf{Description} & \textbf{\#Reqs} \\
\midrule
\multirow{6}{*}{General}
& Theme    & Theme \& Appearance settings & 5 \\

& Navigation & Interaction and navigation & 13 \\

& Content     & Content management & 3 \\ 

& Language & Language settings & 2 \\

& Notification     & Notification settings & 1 \\

& Account  & Profile edit and sharing & 2 \\
\midrule
\multirow{2}{*}{\shortstack[l]{Domain-Specific}}
& Health   & Health and diet management & 4 \\
& Creation & Content creation and editing & 4 \\
\midrule
\multicolumn{3}{r}{\textbf{Total}} & \textbf{34} \\
\bottomrule
\end{tabular}
\end{table}

%% file: src/sub/req_detail.tex
\begin{table}[!t]
\centering
\caption{Benchmark Distribution Across Applications and Requirement Categories}
\label{tab:benchmark}
\vspace{-6pt}
\footnotesize
\renewcommand{\arraystretch}{1.45}
\begin{tabular}{>{\centering\arraybackslash}m{2cm}|>{\centering\arraybackslash}m{1.1cm}|>{\centering\arraybackslash}m{1.1cm}|>{\centering\arraybackslash}m{1.0cm}|>{\centering\arraybackslash}m{1.0cm}|>{\centering\arraybackslash}m{1.2cm}|>{\centering\arraybackslash}m{1.0cm}|>{\centering\arraybackslash}m{1.0cm}|>{\centering\arraybackslash}m{1.0cm}|>{\centering\arraybackslash}m{1.1cm}}
\noalign{\hrule height 0.9pt}
\multirow{2}{*}{\diagbox[width=2.3cm,height=0.81cm]
{\textbf{Application}}{\textbf{Category}}}
& \textbf{Theme}
& \textbf{Navigation}
& \textbf{Content}
& \textbf{Language}
& \textbf{Notification}
& \textbf{Account}
& \textbf{Creation}
& \textbf{Health}
& \textbf{Total} \\
& C/F
& C/F
& C/F
& C/F
& C/F
& C/F
& C/F
& C/F
& \textbf{C}/\textbf{F} \\
\noalign{\hrule height 0.85pt}
App A & 4/4 & 7/2 & 1/0 & 2/2 & 1/0 & 1/0 & 0/0 & 0/0 & 16/8 \\
App B & 3/2 & 4/7 & 1/0 & 2/2 & 1/2 & 0/0 & 0/1 & 2/2 & 13/16 \\
App C & 2/2 & 4/4 & 0/1 & 1/1 & 0/0 & 0/0 & 1/1 & 0/0 & 8/9 \\
App D & 2/2 & 6/6 & 2/0 & 2/0 & 0/0 & 1/0 & 0/0 & 0/0 & 13/8 \\
App E & 4/2 & 6/7 & 0/0 & 1/2 & 0/0 & 1/0 & 2/4 & 0/0 & 14/15 \\
App F & 4/4 & 8/3 & 0/0 & 2/0 & 0/0 & 2/1 & 0/0 & 0/0 & 16/8 \\
\noalign{\hrule height 0.7pt}
\textbf{Total} & 19/16 & 35/29 & 4/1 & 10/7 & 2/2 & 5/1 & 3/6 & 2/2 & \textbf{80}/\textbf{64} \\
\noalign{\hrule height 0.9pt}
\end{tabular}
\end{table}

%% file: src/ch05-evaluation.tex

\section{Evaluation Result}

\subsection{Success Rate (RQ1)}

Table~\ref{tab:rq1-by-app} reports the success rates of \tool across individual phases and End-to-End phase, detailed by applications and requirement categories. Overall, \tool successfully completes \todo{65} of the \todo{80} correct instances, achieving an End-to-End success rate of \todo{81.2\%}. The success rates of three individual phases exceed \todo{90.0\%} each, with Phase 1 achieving the highest performance at \todo{96.3\%}. The three phases fail in \todo{3/80, 6/77, and 6/71} cases, respectively. It indicates that \tool can effectively interpret natural language requirements to navigate into the exact entrance states, and that the generated triggering scripts and oracles correctly capture the intended functional semantics specified in natural language.

By category, \tool performs well on common general requirements, such as \textit{Theme} (73.7\%), \textit{Navigation} (80.0\%), \textit{Language} (100\%), \textit{Account} (100\%) and \textit{Health} (100\%), highlighting its practical applicability for testing malleable mobile applications driven by user-desired requirements. In comparison, the performance of \tool drops in Notification, and Creation categories, achieving success rates of \todo{50.0\%} and \todo{33.3\%} , respectively.
Across different apps, the performance is generally consistent, with success rates consistently above \todo{75.0\%}, indicating \tool's generalization across diverse types of applications.

\summary{\tool achieves an End-to-End success rate of 81.2\%, demonstrating the effectiveness of its requirement-driven pipeline for successful functional correctness testing.}

\input{src/sub/rq11}

\subsection{Discrimination Capability (RQ2)}

Table~\ref{tab:rq2-by-app} shows \tool's discrimination ability by recognizing correct implementations from faulty ones. 
Overall, \tool achieves a precision of \todo{90.3\%}, recall of \todo{81.2\%} and specificity of \todo{89.1\%}. It produces 7 false positives and 15 false negatives, effectively rejects the majority (57, 89.1\%) of 64 faulty instances. The higher precision and specificity, compared to a relatively lower recall, indicate that \tool prioritizes rejecting falsely implemented versions over identifying all correctly implemented ones, which is more desirable in automated and complex scenarios.

\todo{At the category level, for \textit{Account}, \textit{Content}, and \textit{Health}, all three metrics reach \todo{100\%}, indicating that \tool can both accurately accept correct implementations and reliably reject faulty ones. In contrast, \textit{Theme} and \textit{Navigation} achieve high specificities of \todo{93.8\%} and \todo{93.1\%}, respectively and precisions of both \todo{93.3\%}, but comparatively lower recalls of \todo{73.7\%} and \todo{80.0\%}. For \textit{Creation} and \textit{Notification}, the gap becomes larger where \textit{Creation} obtains \todo{50.0\%} precision, \todo{33.3\%} recall, and \todo{83.3\%} specificity, while \textit{Notification} shows only \todo{50.0\%} for all three metrics.} It suggests that \tool consistently exhibits a conservative decision-making tendency even in fine-grained, requirement-specific scenarios. At the application level, the metric distributions remain relatively balanced, without significant change toward either false positives or false negatives, indicating stable and consistent discrimination performance across different applications.

\summary{\tool effectively distinguishes correct implementations from faulty ones, with a precision of \todo{90.3\%}, recall of \todo{81.2\%} and specificity of \todo{89.1\%}, maintaining stable disregard of applications and requirement categories.}

\input{src/sub/rq21}

\subsection{Comparison (RQ3)}

Table~\ref{tab:rq3-combined} presents the stage-wise comparison between \tool and the corresponding baselines. Overall, \tool consistently outperforms all baselines across all phases and metrics, indicating its superior capability in both GUI navigation and oracle construction.

In \textbf{Phase 1}, \tool successfully identifies 77 out of 80 correct cases, achieving a success rate of \todo{96.3\%}, while the \navbase succeeds in only 59 cases (\todo{74.3\%}). This result highlights the effectiveness of our state navigation strategy guided by relevance scores with respect to the given requirements.
\todo{Moreover, it also requires fewer interaction steps on successful cases, (\todo{1.54 vs. 1.98 Avg. Steps}, Table \ref{tab:successful_step_distribution}), indicating higher navigation efficiency.} In \textbf{Phase 2}, \tool achieves a precision of \todo{98.7\%}, recall of \todo{92.5\%}, and specificity of \todo{97.0\%}, demonstrating substantial improvements over \textsc{AugmenTest} by 7.3\%, 38.7\%, and 9.1\%, respectively. We analyze the false positives and false negatives of \textsc{AugmenTest} for which \tool produces a correct detection. \todo{Upon inspection, majority of the false negatives can be attributed to incorrectly generated scripts and assertions that are misaligned with the requirement, leading to incorrect rejection of valid implementations.} In \textbf{Phase 3}, \tool achieves a precision of \todo{91.3\%}, recall of \todo{91.3\%}, and specificity of \todo{72.0\%}, representing consistent improvements of \todo{9.2\%}, \todo{33.8\%}, and \todo{12.0\%} over \textsc{AugmenTest}, respectively. We analyze the false positives and false negatives of \textsc{AugmenTest} for which \tool produces a correct detection. \todo{Meanwhile, the higher Specificity shows that \tool is more effective at rejecting faulty post-condition behaviors, whereas the baseline more frequently accepts incorrect outcomes as valid, indicating weaker discrimination of erroneous results.}

\summary{Compared with the baselines, \tool consistently achieves higher effectiveness across all phases, demonstrating the effectiveness of its intrinsic design and component modules.}

\input{src/sub/rq31}

\subsection{Overhead Evaluation (RQ4)}

We evaluate the time overhead of \tool. Overall, \tool completes the end-to-end pipeline in an average of 246 seconds per instance. Phase 1 and Phase 2 are time-consuming, taking 117 seconds and 101seconds due to the iterative navigation process, while Phase 3 is more efficient, averaging 57seconds.

\subsection{Parameter Sensitivity (RQ5)}

\input{src/sub/rq5}

Table \ref{tab:sensitivity} shows the results of the parameter analysis with varying numbers of candidate $k$. The success rate of \tool increases as $k$ rises from 1 to 3 and remains stable at $k = 4$. To balance effectiveness and efficiency, we set $k = 3$ as the default.

\subsection{Threats and Limitations}

\textbf{Threats.} First, the dataset of apps and user requirements used in our evaluation is limited in scale and diversity, which may affect the generalizability of the results. Nevertheless, the selected apps and requirements have already cover a range of common functionalities and interaction patterns, providing a representative evaluation scenario. Second, our evaluation relies on a single LLM for code generation and relevance scoring and different LLMs may yield varying performance. However, our design is LLM-agnostic and can be adapted to alternative models with minimal modifications.

\textbf{Limitations.} First, \tool is currently designed for Android applications and may not directly generalize to other mobile platforms. However, the underlying methodology of requirement-driven GUI navigation and oracle generation is platform-independent and could be extended to iOS with minimal engineering effort. Second, \tool assumes user requirements are clearly specified and unambiguous, whereas vague or contradictory requirements remains an open challenge and may reduce the accuracy of automated test generation. In the future, we plan to integrate preprocessing techniques to clean and interpret low-quality or unclear user requirements.

%% file: src/sub/rq11.tex
\begin{table}[!t]
\centering
\footnotesize
\caption{Results of \tool's Success Rate}
\vspace{-10pt}
\label{tab:rq1-by-app}
\setlength{\tabcolsep}{5pt}
\renewcommand{\arraystretch}{1.12}
\begin{tabular}{
>{\centering\arraybackslash}p{1.2cm}
>{\raggedright\arraybackslash}p{1.5cm}
>{\centering\arraybackslash}p{1.2cm}
>{\centering\arraybackslash}p{1.2cm}
>{\centering\arraybackslash}p{1.2cm}
>{\centering\arraybackslash}p{1.2cm}
>{\centering\arraybackslash}p{1.7cm}
}

\toprule
\textbf{Type} & \textbf{Category} & \textbf{\#Correct} & \textbf{Phase 1} & \textbf{Phase 2} & \textbf{Phase 3} & \textbf{End-to-End} \\
\midrule
\multirow{6}{*}{App.} & 
App A  & 16 & 15/16 & 13/15 & 12/13 & 12/16 \\
& App B  & 13 & 13/13 & 12/13 & 12/12 & 12/13 \\
& App C  &  8 & 8/8   & 6/8   & 6/6   & 6/8 \\
& App D  & 13 & 12/13 & 11/12 & 10/11 & 10/13 \\
& App E  & 14 & 14/14 & 14/14 & 12/14 & 12/14 \\
& App F  & 16 & 15/16 & 15/15 & 13/15 & 13/16 \\
\midrule
\multirow{8}{*}{Req.}
& Theme        & 19 & 19/19 & 16/19 & 14/16 & 14/19 \\
& Navigation   & 35 & 33/35 & 31/33 & 28/31 & 28/35 \\
& Content      &  4 &   4/4 &   4/4 &   4/4 &   4/4 \\
& Language     & 10 & 10/10 & 10/10 & 10/10 & 10/10 \\
& Notification &  2 &   1/2 &   1/1 &   1/1 &   1/2 \\
& Account      &  5 &   5/5 &   5/5 &   5/5 &   5/5 \\
& Creation     &  3 &   3/3 &   2/3 &   1/2 &   1/3 \\
& Health       &  2 &   2/2 &   2/2 &   2/2 &   2/2 \\
\midrule
\textbf{Overall} & - & \textbf{80} 
& \begin{tabular}[c]{@{}c@{}}\textbf{77 / 80}\\ \textbf{(96.3\%)}\end{tabular} 
& \begin{tabular}[c]{@{}c@{}}\textbf{71 / 77}\\ \textbf{(92.2\%)}\end{tabular} 
& \begin{tabular}[c]{@{}c@{}}\textbf{65 / 71}\\ \textbf{(91.5\%)}\end{tabular} 
& \begin{tabular}[c]{@{}c@{}}\textbf{65 / 80}\\ \textbf{(81.2\%)}\end{tabular} \\
\bottomrule
\end{tabular}
\end{table}

%% file: src/sub/rq21.tex
\begin{table}[!t]
\centering
\footnotesize
\caption{Results of \tool's Discrimination Capability}
\label{tab:rq2-by-app}
\vspace{-10pt}
\begin{tabular}{
>{\centering\arraybackslash}p{1.1cm}
>{\raggedright\arraybackslash}p{1.3cm}
>{\centering\arraybackslash}p{0.8cm}
>{\centering\arraybackslash}p{0.8cm}
>{\centering\arraybackslash}p{0.8cm}
>{\centering\arraybackslash}p{0.8cm}
>{\centering\arraybackslash}p{1.2cm}
>{\centering\arraybackslash}p{1.2cm}
>{\centering\arraybackslash}p{1.2cm}
}
\toprule
\textbf{Type} & \textbf{Category}& \textbf{TP} & \textbf{FN} & \textbf{TN} & \textbf{FP} & \textbf{Precision} & \textbf{Recall} & \textbf{Specificity} \\
\midrule
\multirow{6}{*}{App.}
& App A & 12 & 4 &  6 & 2 & 85.7 & 75.0 & 75.0 \\
& App B & 12 & 1 & 14 & 2 & 85.7 & 92.3 & 87.5 \\
& App C &  6 & 2 &  9 & 0 & 100  & 75.0 & 100  \\
& App D & 10 & 3 &  7 & 1 & 90.9 & 76.9 & 87.5 \\
& App E & 12 & 2 & 13 & 2 & 85.7 & 85.7 & 86.7 \\
& App F & 13 & 3 &  8 & 0 & 100  & 81.2 & 100 \\
\midrule
\multirow{8}{*}{Req.}
& Theme        & 14 &  5 & 15 & 1 & 93.3 & 73.7 & 93.8  \\
& Navigation   & 28 &  7 & 27 & 2 & 93.3 & 80.0 & 93.1  \\
& Content      &  4 &  0 &  1 & 0 & 100  & 100  & 100   \\
& Language     & 10 &  0 &  5 & 2 & 83.3 & 100  & 71.4  \\
& Notification &  1 &  1 &  1 & 1 & 50.0 & 50.0 & 50.0  \\
& Account      &  5 &  0 &  1 & 0 & 100  & 100  & 100   \\
& Creation     &  1 &  2 &  5 & 1 & 50.0 & 33.3 & 83.3  \\
& Health       &  2 &  0 &  2 & 0 & 100  & 100  & 100   \\
\midrule
\textbf{Overall} & - & \textbf{65} & \textbf{15} & \textbf{57} & \textbf{7} & \textbf{90.3} & \textbf{81.2} & \textbf{89.1} \\
\bottomrule
\end{tabular}
\end{table}

%% file: src/sub/rq31.tex
\begin{table}[t]
\centering
\footnotesize
\setlength{\tabcolsep}{2pt}
\caption{Results of Comparison Evaluation between \tool and Baselines}
\vspace{-10pt}
\label{tab:rq3-combined}
\begin{tabular}{@{}
p{2.0cm}
>{\centering\arraybackslash}p{2.2cm}
@{\hspace{17pt}}
>{\centering\arraybackslash}p{1.4cm}
@{\hspace{2pt}}
>{\centering\arraybackslash}p{1.05cm}
@{\hspace{2pt}}
>{\centering\arraybackslash}p{1.4cm}
@{\hspace{17pt}}
>{\centering\arraybackslash}p{1.4cm}
@{\hspace{2pt}}
>{\centering\arraybackslash}p{1.05cm}
@{\hspace{2pt}}
>{\centering\arraybackslash}p{1.6cm}
@{}}
\toprule
\multirow{2}{*}{\textbf{Method}} 
& \textbf{Phase 1} & \multicolumn{3}{c}{\textbf{Phase 2}} & \multicolumn{3}{c}{\textbf{Phase 3}} \\
\cmidrule(lr{17pt}){2-2} \cmidrule(l{1pt}r{17pt}){3-5} \cmidrule(l{1pt}r{1pt}){6-8}
& \textbf{Success Rate} & \textbf{Precision} & \textbf{Recall} & \textbf{Specificity} & \textbf{Precision} & \textbf{Recall} & \textbf{Specificity} \\
\midrule
\todo{\navbase}     
& 59/80 (74.3\%) & -- & -- & -- & -- & -- & -- \\
\todo{\oraclebase}     
& -- & 91.4\% & 53.8\% & 87.9\% & 82.1\% & 57.5\% & 60.0\% \\
\midrule
\textbf{\tool} 
& \textbf{77/80 (96.3\%)} & \textbf{98.7\%} & \textbf{92.5\%} & \textbf{97.0\%} & \textbf{91.3\%} & \textbf{91.3\%} & \textbf{72.0\%} \\
\bottomrule
\end{tabular}
\end{table}

\begin{table}[t]
\centering
\footnotesize
\setlength{\tabcolsep}{2pt}
\caption{Distribution of Successful Steps for \navbase and \tool in Phase1.}
\label{tab:successful_step_distribution}
\vspace{-10pt}
\begin{tabular}{@{}
l
>{\centering\arraybackslash}p{1.8cm}
>{\centering\arraybackslash}p{0.9cm}
>{\centering\arraybackslash}p{1.1cm}
>{\centering\arraybackslash}p{1.1cm}
>{\centering\arraybackslash}p{1.1cm}
>{\centering\arraybackslash}p{1.1cm}
>{\centering\arraybackslash}p{1.1cm}
>{\centering\arraybackslash}p{1.1cm}
>{\centering\arraybackslash}p{0.9cm}
>{\centering\arraybackslash}p{0.9cm}
@{}}
\toprule
\multirow{2}{*}{\textbf{Method}} 
& \multirow{2}{*}{\textbf{Succ.}} 
& \multicolumn{7}{c}{\textbf{No. of Successful Steps ($n$/\%)}} 
& \multirow{2}{*}{\textbf{Avg}} 
& \multirow{2}{*}{\textbf{Med}}\\
\cmidrule(lr){3-9}
&& \textbf{0} & \textbf{1} & \textbf{2} & \textbf{3} & \textbf{4} & \textbf{6} & \textbf{9} &  &  \\
\midrule
\navbase & 59/80 (74.3\%) & 9/15.3\% & 14/23.3\% & 20/33.9\% & 8/13.6\% & 6/10.2\% & 1/1.7\% & 1/1.7\% & 1.98 & 2 \\
\textbf{\tool} & \textbf{77/80 (96.3\%)} & \textbf{13/16.9\%} & \textbf{24/31.2\%} & \textbf{25/32.5\%} & \textbf{11/14.3\%} & \textbf{3/3.9\%} & \textbf{0/0.0\%} & \textbf{0/0.0\%} & \textbf{1.54} & \textbf{1} \\
\bottomrule
\end{tabular}
\end{table}

%% file: src/sub/rq5.tex
\begin{table}[t]
\centering
\footnotesize
\setlength{\tabcolsep}{7pt}
\caption{Parameter Analysis of using Different Candidate Numbers in Phase 1}
\label{tab:sensitivity}
\vspace{-10pt}
\begin{tabular}{@{}lccccc@{}}
\toprule
\textbf{Parameter Value} & \textbf{Candidate=1} & \textbf{Candidate=2} & \textbf{Candidate=3} & \textbf{Candidate=4} \\
\midrule
\textbf{Success number} & 71/80 & 76/80 & \textbf{77/80} & 77/80 \\
\textbf{Success rate} & 88.75\% & 95\% & \textbf{96.25\%} & 96.25\% \\
\bottomrule
\end{tabular}
\end{table}

%% file: src/ch06-related.tex

\section{Related Work}

\textbf{Large Language Models for Code Generation.} Recent advances in LLMs have significantly improved code generation capabilities, though issues such as hallucination remain~\cite{zhang2025llm}. Benchmarks like HumanEval~\cite{chen2021codex}, CrossCodeEval~\cite{ding2023crosscodeeval}, and RepoMasterEval~\cite{wu2025repomastereval} have been proposed to evaluate LLMs from function-level to repository-level settings. At the repository level code generation, many approaches focus on retrieval problems. For example, Zhang et al.~\cite{zhang2023repocoder} introduces RepoCoder, which performs iterative retrieval to progressively incorporate repository context into the generation process. Cheng et al.~\cite{cheng2024dataflow} leverages dataflow analysis to retrieve semantically relevant code snippets, while Liu et al.~\cite{liu2024graphcoder} constructs a code context graph and performs coarse-to-fine retrieval to capture repositories' structural dependencies. In addition, Wu et al.~\cite{wu2024repoformer} proposes selective retrieval to avoid unnecessary or noisy retrieved contexts. Zhang et al.~\cite{zhang2025hierarchical} investigates redundant repository context reduction through hierarchical pruning strategies. Other methods explore agent-based retrieval~\cite{zhang2024codeagent} and reinforcement learning optimization~\cite{wang2025rlcoder}. Another line of work improves input quality. Techniques include inferring missing requirements~\cite{han2024archcode}, interactive clarification~\cite{mu2024clarifygpt}, and structured prompting~\cite{ma2025should, zhang2025little}. Fakhoury et al.~\cite{fakhoury2024llm, fakhoury2024exploring} leverage test-driven intent clarification for informal natural language specifications and Dong et al.~\cite{dong2025contested} incorporates user feedback to urther enhance generation quality.

\textbf{Large Language Models for Testing.} LLMs enable automated unit test generation, though naive application often underperforms~\cite{schafer2023empirical}. To improve coverage, Wang et al.\cite{wang2024hits} generate tests via decomposed focal methods, Ryan et al.\cite{ryan2024code} use multi-stage prompting, Cheng et al.\cite{cheng2025rug} apply semantic-aware context and fuzzing for Rust, and Nan et al.\cite{nan2025test} leverage explicit test inputs and mock behaviors. LLMs also assist test maintenance and repair~\cite{zhang2025unit, rahman2025utfix, chen2024chatunitest}, and real-time testing~\cite{harman2025harden}. Beyond test generation, LLMs have been applied to produce test oracles~\cite{molina2025test}, unit assertions~\cite{zhang2025exploring, molinelli2025llms}, and Verilog assertions~\cite{pulavarthi2025assertionbench, pulavarthi2025llms}.  Shin et al.~\cite{shin2024assessing} highlight the limitations of textual similarity metrics. Motivated by the limitation, approaches incorporate analysis and feedback, such as fine-tuning and customized prompting~\cite{hossain2025togll}, static and dynamic feedback~\cite{hayet2024chatassert}, functional coverage~\cite{wang2026iterative}, dedicated training~\cite{zhou2025insights}
Recently, based on the potential of the LLMs, oracles are also generated from failure reports~\cite{baral2024automating, johnson2025generating} or natural language specifications~\cite{hossain2025doc2oracll, khandaker2025augmentest, endres2024can, kande2024security}.

\textbf{Malleable Software.} Malleable software emphasizes dynamic and evolving behavior, interpreted in various forms across research domains. Gobert et al.~\cite{gobert2023lorgnette} present a framework that allows programmers to augment code editors with customizable code projections that represent programs in alternative visual forms (i.e., tables, graphs, or forms). However, their visual forms are rigidly defined. More advanced approaches generate malleable UIs from user tasks or requirements. {Cao et al.~\cite{cao2025generative} leverages LLMs to implement a generative and malleable UI approach, enabling end-users to iteratively customize interfaces through natural language.} Chen et al.~\cite{chen2025taskartisan} also present TaskArtisan, which translates multimodal input into structured specifications and synthesizes interactive widgets. Apart from general user interface malleability, there are software malleability realized to specific problems. Aveni et al.~\cite{aveni2025generative} introduce Ply, which generates behaviors from natural language with layered abstractions. Zamfirescu et al.\cite{zamfirescu2025beyond} show that a customized IDE helps developers explore problem and solution spaces. Moreover, Martin~\cite{martin2023efficient} and MaM~\cite{martin2024mam} implement process malleability in MPI applications for efficient dynamic redistribution.

%% file: src/ch07-conclusion.tex

\section{Conclusion}

We envision a paradigm where end-users can directly specify their requirements, enabling per-user malleable mobile applications. To support this vision, we propose \tool, a requirement-driven GUI test generation framework that efficiently navigates the UI, triggers user-specified functionalities, and constructs LLM-guided oracles to validate correctness. We build a comprehensive benchmark and perform extensive experiments spanning six popular mobile applications.
Our study highlights the feasibility and potential of end-user-driven software development, paving the way for the malleable software, which is more adaptable, personalized, and user-centered.